\documentclass[pre,aps,twocolumn,showpacs,superscriptaddress,floatfix]{revtex4}

\usepackage{amssymb}
\usepackage{graphicx}
\usepackage{amsmath}
\usepackage{latexsym}
\usepackage{verbatim}
\usepackage{bm}

\begin{document}

\title{Map equation for link community}

\author{Youngdo Kim}
\affiliation{
Department of Physics, Korea Advanced Institute of Science and Technology, Daejeon 305-701, Korea.}

\author{Hawoong Jeong}
\email{hjeong@kaist.edu}
\affiliation{
Department of Physics, Korea Advanced Institute of Science and Technology, Daejeon 305-701, Korea.}
\affiliation{
Institute for the BioCentury, Korea Advanced Institute of Science and Technology, Daejeon 305-701, Korea.}
\date{\today}% It is always \today, today,
             %  but any date may be explicitly specified

\begin{abstract}
Community structure exists in many real-world networks and has been reported being related to several functional properties of the networks. The conventional approach was partitioning nodes into communities, while some recent studies start partitioning links instead of nodes to find overlapping communities of nodes efficiently. We extended the map equation method, which was originally developed for node communities, to find link communities in networks. This method is tested on various kinds of networks and compared with the metadata of the networks, and the results show that our method can identify the overlapping role of nodes effectively. The advantage of this method is that the node community scheme and link community scheme can be compared quantitatively by measuring the unknown information left in the networks besides the community structure. It can be used to decide quantitatively whether or not the link community scheme should be used instead of the node community scheme. Furthermore, this method can be easily extended to the directed and weighted networks since it is based on the random walk.
\end{abstract}


\maketitle

\section{introduction}
Complex networks have been widely used to represent the systems composed of connected objects, and many system-wide behaviors, which have emerged from the pattern of connections, have been successfully explained with the help of this simple model~\cite{Newman2003a,Albert2002}. The rising popularity of complex network through several disciplines---including statistical physics, computer science, computational biology, sociology, etc.---rests on many reasons; a major one is that many large scale networks have become available due to the advance in information technology. The advantage of the large-scale networks is that many meaningful statistical properties can be studied accurately, for example, degree distribution, clustering coefficient, assortativity, and motif profiles. However, the big size of the networks also brings disadvantages. When the network is small, it is very easy to visualize the network, and the organization structure of the network can be perceived intuitively. Instead, when the size becomes large, a comprehensive understanding of the structure could no longer be gained directly, and some quantitative analyses are required.

Community detection is one of the efforts devoted to the quantitative analysis of the organization structure. In many real-world networks, the nodes are connected neither regularly nor completely randomly. Instead, some nodes are densely inter-connected to form the communities, while these communities are loosely connected, relatively. This kind of network structure, which is usually referred as the community structure, is closely related to many dynamic processes on the network~\cite{Colizza2006a,Guimera2006}. Therefore, detecting the community structure has become one of the most important problems in the network research and many methods have been proposed to solve the problem efficiently~\cite{Fortunato2009}. The map equation method~\cite{Rosvall2007}, also known as Infomap method, has been considered one of the best performing methods~\cite{YYAhn2009, Fortunato2009}. This method is based on the Minimum Description Length (MDL) principle~\cite{Rissanen2004}, according to which any regularity in the data can be used to compress the length of the data. Therefore, by considering the community structure as the regularity of the network and the path of the random walk on the network as the data to compress, the community structure can be detected during the compression of the path description. This is the main idea of the map equation method and it will be explained in detail in Sec.~\ref{sec:me_nodecom}

While most previous researches for community detection have focused on the community of nodes, some recent researches have started switching attention to community of links~\cite{YYAhn2009, Evans2009a} and even cliques~\cite{Evans2010}.  From the theoretical point of view, the community of link could be more intuitive than the community of node in some real-world networks, because the link is more likely to have a unique identity while the node tends to have multiple identities. For example, most individuals in the society belong to multiple communities such as families, friends, and co-workers while the link between a pair of individuals usually exists for a dominant reason. From the practical point of view, overlapping communities of nodes, which is another attractive topic of community detection~\cite{Nepusz2008,Gregory2011,Kovacs2010a,Esquivel2011} could be detected as a byproduct because the links connected to a single node could belong to different link communities and consequently the node could be assigned to multiple communities of links. But exclusive partitioning of links is not always accurate and this problem is discussed in Sec.~\ref{sec:RE0}. The clique community is going further in this direction since a link is a clique of two nodes.

In this paper, we propose a modified version of the map equation method, which can be used to detect link communities under the MDL principle. In Sec.~\ref{sec:me_nodecom}, a brief review of original map equation is presented and the modified version of the map equation method is introduced in the following Sec.~\ref{sec:me_linkcom}. The best way to check the performance of a community detecting method is to compare the community result with the metadata available. We apply our method to several networks with rich metadata information, and the results are quantitatively compared with other methods for community detection  in Sec.~\ref{sec:RE1}. An important advantage of our method is that the results of link community and node community can be quantitatively compared. In Sec.~\ref{sec:RE2}, a model network is proposed to verify this property and the comparison is done in some real-world networks to show which partitioning scheme---link community or node community---can depict the organization structure of these networks more properly. 

For the simplicity of derivation, only the binary and undirected network is considered in this paper. The extension to weighted and/or directed networks is briefly discussed at the end of Sec.~\ref{sec:me_linkcom}.

\section{The Map Equation for Node Community}
\label{sec:me_nodecom}
The most general definition of the community is that a community is a group of nodes that are densely inter-connected. Meanwhile, from the viewpoint of information propagation, another definition can be proposed: A community is a group of nodes in which the information is more likely to be trapped rather than spread out. Considering that the random walk is the most fundamental model of information propagation, community structure can be detected by finding the local structure that traps the random walker. Some recent studies~\cite{ Delvenne2008,YKim2009} have showed that the modularity~\cite{Newman2006}, which is a quality function used to find the communities as a group of densely connected nodes, can also be interpreted by the random walk and some disadvantages of the modularity can be easily resolved in this alternative approach. 

The map equation method~\cite{Rosvall2007} detects communities by the information-propagation-based definition, under the philosophy of Minimum Description Length (MDL) principle~\cite{Rissanen2004}. The basic idea of the MDL principle is that any regularity in the data can be used to compress the length of the data. If we can find a way to encode the path of random walk on the network and consider the community structure as the regularity in the network, community structure can be detected by finding the partition that gives the minimum description length of the path. In the map equation method, the encoding rule for the path description can be described as follows.

To uniquely describe the path of a random walk on the network, the simplest way would be assigning a distinguishable code to each node in order to avoid the ambiguity, and the description length would become shorter when the more frequently visited nodes are given shorter code and less frequently nodes given a relatively longer code, which is the method known as the Huffman coding~\cite{Huffman1952}. However, assigning a unique code to each node in the network could be very inefficient if the network size is large, and the movement of the random walker is frequently trapped in a small area---the community of nodes. A better strategy would be dividing the nodes into communities and using the codebook of two levels: The first level code describes the community that a node belongs to, and the second level code distinguishes a specific node from other nodes in the same community. In this strategy, a community (first level) code should be recorded in the path description when and only when the random walker enters the new community from other communities, and the random walks that is taking place within the community can be uniquely described by recording only the second level code. Additionally, an exit code should be assigned to each community, and it should be recorded when the random walker is exiting a community, so that the first level code and the second level codes can be distinguished. The costs of using the two-level codes would be fully compensated if the community structure is significant and it is well detected, because in this case the second level codes would become much shorter, and the first level codes of communities would not be frequently used, consequently reducing the total length of the path description. Therefore, the best partition of the network would be the partition that minimizes the average description length of the path of the random walk under the coding strategy described above.

Once the community partition $\bm{M}$ is decided, the probability of each code being used can be easily calculated and the map equation  $L_\textrm{nodecom}(\bm{M})$, which is defined as the theoretical minimum of average description length, can be given by the Shannon's source coding theorem~\cite{Shannon2001} as
\begin{equation}
L_\textrm{nodecom}(\bm{M}) = q_{\curvearrowright}H(Q)+\sum_{i=1}^{C}{p^{i}_{\circlearrowright}}H(P^i),
\label{eq:MENode}
\end{equation}
where $i$ is the index of community, $\alpha$ is the index of node, and $C$ is the number of communities; $q_{\curvearrowright} \equiv \sum_{i=1}^{C} q^{i}_{\curvearrowright}$ is the total probability of using the first level codebook where $q^{i}_{\curvearrowright}$ is the probability of using the first level code for community $i$; $p^{i}_{\circlearrowright} \equiv q^{i}_{\curvearrowright}+\sum_{\alpha \in i}{p_\alpha}$ is the probability of using the second level codebook and the exit code for community $i$; and $p_\alpha$ is the probability of node $\alpha$ being visited, which is equal to the probability of using the second level code for node $\alpha$. $H(Q)$ is the average description length contributed by the first level codebook:
\begin{equation}
H(Q)=-\sum_{i=1}^{C}{\frac{q^{i}_{\curvearrowright}}{q_{\curvearrowright}}\log(\frac{q^{i}_{\curvearrowright}}{q_{\curvearrowright}})},
\label{eq:Level1Node}
\end{equation}
while $H(P^i)$ is the description length contributed by the second level codebook for community $i$:
\begin{eqnarray}
\label{eq:Level2Node}
H(P^i) =  -\frac{q^{i}_{\curvearrowright}}{q^{i}_{\curvearrowright}+\sum_{\alpha \in i}{p^i_\alpha}}\log({\frac{q^{i}_{\curvearrowright}}{q^{i}_{\curvearrowright}+\sum_{\alpha \in i}{p^i_\alpha}}}) \nonumber \\
 -\sum_{\alpha \in i}{\frac{p^i_\alpha}{q^{i}_{\curvearrowright}+\sum_{\beta \in i}{p^i_\beta}}\log({\frac{p^i_\alpha}{q^{i}_{\curvearrowright}+\sum_{\beta \in i}{p^i_\beta}}})},
\end{eqnarray}
where $p^i_\alpha$ is equal to $p_\alpha$ when node $\alpha$ belongs to community $i$, otherwise zero. The probability $q^{i}_{\curvearrowright}$ is included in Eq.~(\ref{eq:Level2Node}) to represent the contribution of exit codes for community $i$, and it can be computed from the following equation once the community structure $\bm{M}$ is given:
\begin{eqnarray}
q^{i}_{\curvearrowright} = \sum_{\alpha \in i}\sum_{\beta \notin i}{p_{\alpha} \frac{A_{\alpha \beta}}{k_\alpha}},
\end{eqnarray}
where $A_{\alpha \beta}$ is the element of the adjacency matrix, and it equals one if there is a link between node $\alpha$ and $\beta$, otherwise zero; $k_{\alpha} \equiv \sum_{\beta}{A_{\alpha \beta}}$ is the degree of node $\alpha$. The description length is measured in bits if the logarithm is taken with base $2$ in the equations above.

The community structure can be detected by finding the partition of nodes that minimizes the map equation $L_\textrm{nodecom}(\bm{M})$ in Eq.~(\ref{eq:MENode}), just like other community detection methods based on maximization (or minimization) of the quality function. For example, many algorithms developed to maximize the modularity~\cite{Newman2006} can be directly used to minimize the map equation by replacing only the definition of quality function in the algorithms.

\section{The Map Equation for Link Community}
\label{sec:me_linkcom}

\begin{figure}
\includegraphics[width=0.9\columnwidth]{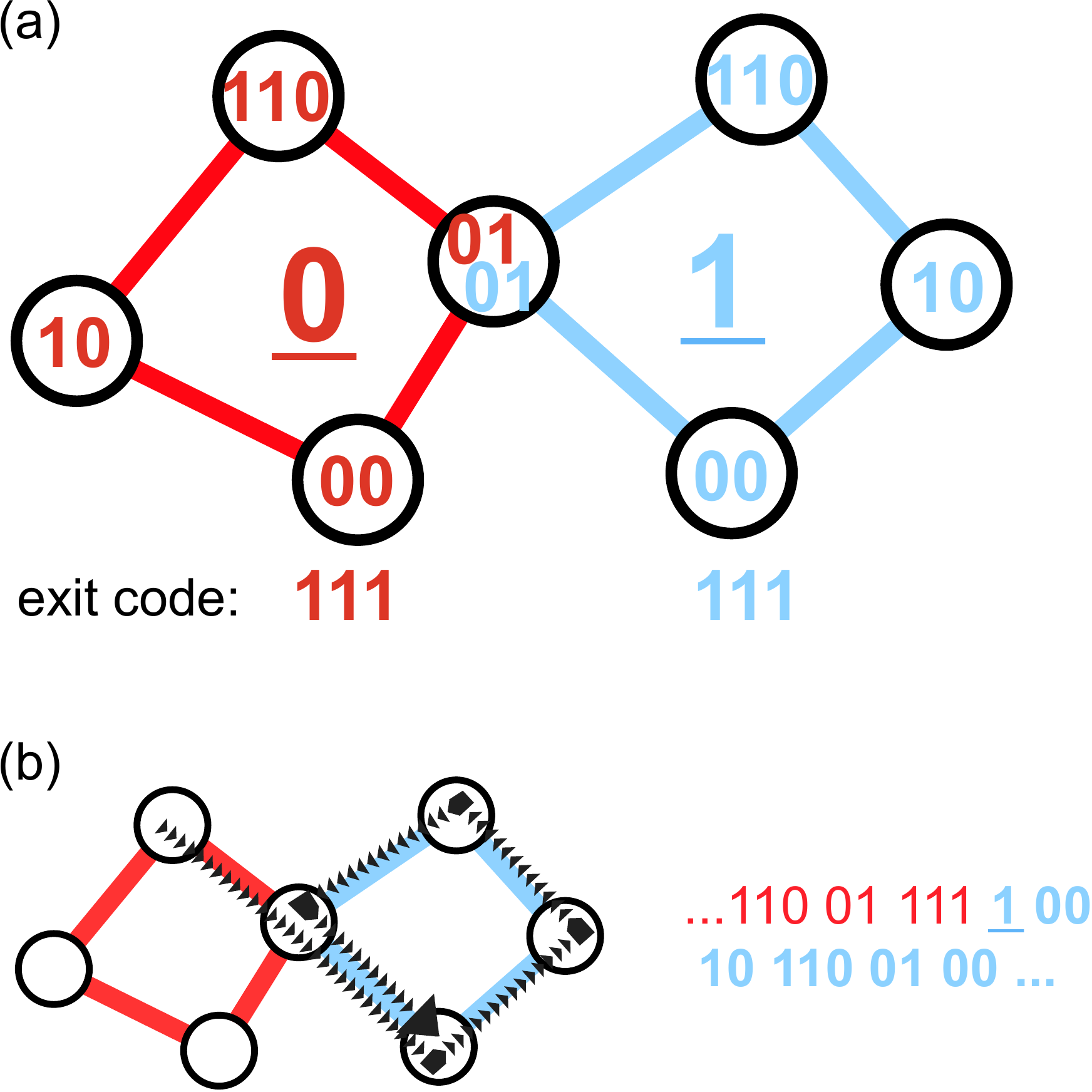}
\caption{\label{fig:code} (Color online) The encoding rule for the random walk path description in our method. (a) The links are divided into two communities: the red (dark gray) one and the blue (light gray) one, and the underlined first level codes are assigned to each link community. The second-level codes are assigned to the nodes, and the node in the center is given two second-level codes because it belongs to both communities. (b) An example of the random walk path is depicted on the left side, and the description of the path is given on the right side. The underlined first-level code is recorded only when the random walker is moving across the communities, and it is omitted when the random walker is moving within the community.}
\end{figure}

Although various kinds of methods~\cite{Fortunato2009} have been developed to find the communities, most of them are limited to the community of nodes. Some recent studies~\cite{YYAhn2009,Evans2009a}, in which the link community is studied instead of the node community, showed that if the focus is alternated from the nodes to the links, a better description of the community structure could be found. In many real-world networks, a node could belong to several communities at the same time, and this fact makes the node community scheme fail to describe the organization structure of the system properly. For example, a person can belong to several social groups at the same time and an interdisciplinary research can belong to several scientific fields. Meanwhile, a link between a pair of nodes usually exists for a dominant reason, and the overlaps of communities over links would be less likely to happen compared to the overlaps of communities over nodes. The immediate advantage of the link community is that it can be used to detect overlapping communities of nodes, which is another active field of community identification~\cite{Nepusz2008,Gregory2011}. Although a link belongs only to a specific community when the links are partitioned into communities, a node could belong to multiple communities because the links connected to the node could belong to different communities (i.e., the link communities are overlapping over the nodes). A similar discussion can be applied to the cliques~\cite{Evans2009a}, which are the subnetworks of fully connected nodes, and the link community can be considered as a special case of clique community since a link is a clique composed of two nodes.

In this section, we propose a modified version of the map equation that can be used to find the communities of links. Since the original map equation can only be applied for node community, the encoding rule for the path of random walk needs to be modified. As illustrated in Figure~\ref{fig:code}, the first step of this modification is to let the partition $\bm{M}$ describe the link community instead of node community. The links are partitioned into communities, and the first level code is assigned to each link community. Meanwhile, the second level codes are still assigned to the nodes. The advantage of this encoding rule will be discussed later in Sec.~\ref{sec:RE2}.  Since some nodes could belong to multiple communities in this case, each of these overlapping nodes would be given multiple second level codes, as many as the number of communities the node belongs to. Once the first- and second-level codes are assigned according to the community structure we assume, the path description is given as: (i) at each step, the random walker is moving from the source node to the target node, which means the random walker is moving over a selected link that connects the source and target nodes; (ii) if the link bypassing at current step belongs to a different community compared to the community that the link of previous step belongs to, the first level code for community is recorded before recording the second level code for the target node; (iii) if the links of the current step and the previous step belong to the same community, the first level code would be omitted and only the second level code for the target node is recorded; (iv) additionally, an exit code should be inserted before each first level code in order to distinguish the first level codes from the second level codes. 

The nodes that belong to multiple communities have multiple second level codes and this redundancy is likely to increase the length of the path description. However, if the link community is more significant than the node community (i.e., many nodes belong to multiple communities), the redundancy can be compensated by reducing the frequency of using first level codes especially when the random walker visits the overlapping nodes and move back to the previous community.

Once the encoding rule is given as above, we can get the map equation for link community if we know about the probability of using each code, and this computation of each probability can be easily done with the help of LinkRank~\cite{YKim2009}. LinkRank $r_{\alpha\beta}$, which is the probability of the link $\alpha \rightarrow \beta$ being visited by the random walker in the stationary state, is a constant value equal to $1/2M$ in the undirected  binary networks, where $M$ is the number of links in the network. We use $r^i_{\alpha\beta}$ to represent the community partition $\bm{M}$: $r^i_{\alpha\beta}$ is equal to $r_{\alpha\beta}$ if the link between nodes $\alpha$ and $\beta$ belongs to community $i$, otherwise zero. Given the probability of visiting each link, the probability of using a second level code for node $\alpha$  in the community $i$ is 
\begin{equation}
p^{i}_{\alpha}=\sum_{\beta}{r^{i}_{\beta\alpha}}, 
\label{eq:pi}
\end{equation}
and the probability of using the first level code for community $i$ is $q^{i}_{\curvearrowright}=\sum_{\alpha}{q^{i}_{\alpha\curvearrowright}}$, where
\begin{equation}
q^{i}_{\alpha\curvearrowright}=p^{i}_{\alpha}(1-\frac{\sum_{\beta}{r^{i}_{\alpha\beta}}}{p_\alpha}),
\label{eq:qi}
\end{equation}
is the probability that the first level code being used after visiting node $\alpha$. Here $p_\alpha$ is the probability of visiting node $\alpha$ and it satisfies $p_\alpha = \sum_{i}{p^i_\alpha} = {k_\alpha} /{2M}$, where $k_\alpha$ is the degree of node $\alpha$. 

Finally, the map equation for the link community can be given as 
\begin{equation}
L_\textrm{linkcom}(\bm{M}) = q_{\curvearrowright}H(Q)+\sum_{i=1}^C{p^{i}_{\circlearrowright}}H(P^i),
\label{eq:MElinkcom}
\end{equation}
where $q_{\curvearrowright} = \sum_{\alpha,i}{q^{i}_{\alpha\curvearrowright}}$ is the total probability of using first level codes, and $p^{i}_{\circlearrowright}= q^{i}_{\curvearrowright}+\sum_{\alpha}{p^{i}_{\alpha}}$ is the total probability of using second level codes and the exit codes. $H(Q)$ is the contribution of first level codes to the average description length, and it can be computed by
\begin{equation}
H(Q)=-\sum_{i=1}^{C}{\frac{q^{i}_{\curvearrowright}}{q_{\curvearrowright}}\log(\frac{q^{i}_{\curvearrowright}}{q_{\curvearrowright}})}.
\end{equation}
 Similarly, $H(P^i)$ is the contribution of second level codes in community $i$ to the average description length, and it can be computed from the following equation
\begin{eqnarray}
H(P^i) =  -\frac{q^{i}_{\curvearrowright}}{p^{i}_{\circlearrowright}}\log\frac{q^{i}_{\curvearrowright}}{p^{i}_{\circlearrowright}} 
-\sum_\alpha {\frac{p^i_\alpha}{p^{i}_{\circlearrowright}}\log\frac{p^i_\alpha}{p^i_\circlearrowright}}.
\end{eqnarray}

Now this map equation for link community can be used as the quality function to find link communities, just like other quality functions of community detection. Thus, most of the algorithms developed for other quality functions can also be modified to minimize $L_\textrm{linkcom}(\bm{M}) $ in Eq.~(\ref{eq:MElinkcom}). In this paper, we used a modified version of the algorithm developed by Rosvall and Bergstrom~\cite{Rosvall2009a}, which is an extended version of the Louvain method~\cite{Blondel2008}.  The difference between our optimizing algorithm and the Louvain method is that the links, instead of the nodes, are locally grouped together to find the minimum efficiently.

This method can be easily generalized to weighted networks, in which weight is assigned to each link, and directed networks, in which direction is assigned to each link. In the weighted networks, the LinkRank $r_{\alpha \beta}$ is no longer a constant value, and it is proportional to the weight $w_{\alpha \beta}$ of each link. The remaining processes would just be the same. In the directed networks, the LinkRank $r_{\alpha \beta}$ is a quantity related to the global structure of the network, and it can be computed by following the processes described in Ref.~\cite{YKim2009}. If the directed network is composed of only one strongly connected component (SCC), in which a directed path always exists between any two nodes in the network, the equations in this section can still be directly used. It is important to notice that the sequences of $\alpha$ and $\beta$ in Eqs.~(\ref{eq:pi}) and (\ref{eq:qi}) are different. In the directed networks composed of more than one SCCs, the situation becomes complicated because there would be more than one stationary values for LinkRank. Therefore, the random hopping should be included in the random walk, which is the same as adding all-to-all links of small weight to the network, to ensure the existence of only one stationary value for the LinkRank. Thus, the original network becomes a all-to-all connected network and a link exists between any pair of nodes. This would make the minimization of the map equation computationally expensive, because the number of links to be partitioned would grow significantly. One possible solution is considering the random hopping links only when computing the LinkRank values and then normalize the LinkRank after removing the links generated by random hopping, as previously shown in Ref.~\cite{Rosvall2010a}.

During the submission of this paper, another extension of the map equation for overlapping community is proposed by Esquivel and Rosvall~\cite{Esquivel2011}. It would be an interesting work to compare the performance of these two methods.

\section{A Real-World Network Analysis: The Karate Club Network
\label{sec:RE0}}

\begin{figure}
\includegraphics[width=0.9\columnwidth]{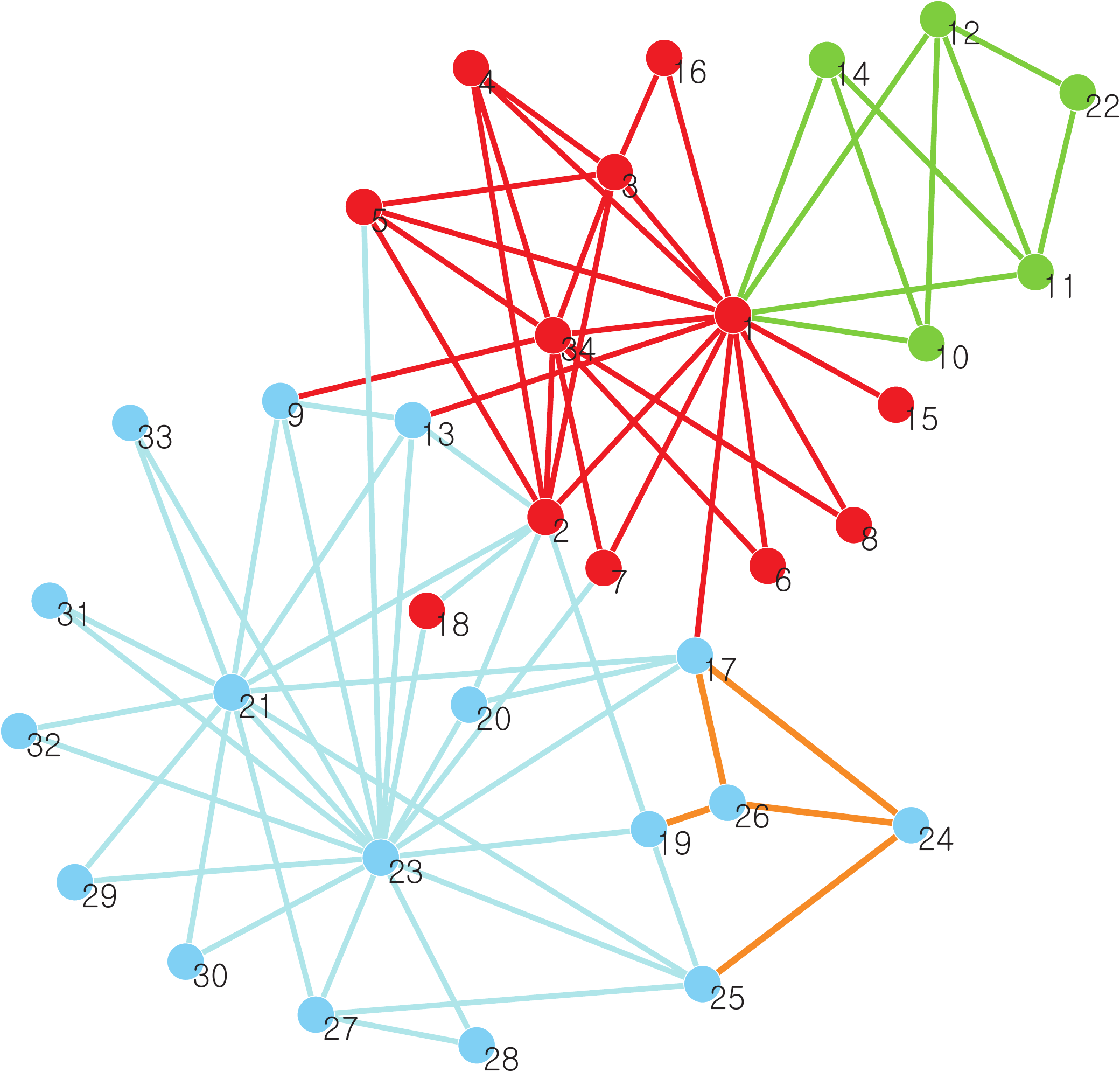}
\caption{\label{fig:karate} (Color online) The communities detected in the karate club network. The color of the nodes indicate three node communities detected by the original map equation method, which is minimizing $L_\textrm{nodecom}$. The color of the links indicate four link communities detected by our method, which is minimizing $L_\textrm{linkcom}$. $L_\textrm{linkcom} = 4.28$ bits and  $L_\textrm{nodecom} = 4.31$ bits, for the community results illustrated in the figure.}
\end{figure}

We applied our method to the famous karate club network~\cite{Zachary1977}, a social network analyzed in most community detection researches, and the result is illustrated in Figure~\ref{fig:karate}. The color of the links indicates the link communities detected by minimizing the map equation for link community, $L_\textrm{linkcom}$, while the color of the nodes indicates the node communities detected by minimizing the map equation for node community, $L_\textrm{nodecom}$. According to the result of node community, some nodes, especially nodes Nos.1 and 2, are categorized in the red (center) community, while a large portion of their neighbors belong to another community. In this karate club society, these nodes should be the members who connect different groups of people together, and their existence would be very important to integration of the whole society. However, the multiple social roles of these nodes are not captured in the node community scheme because the nodes are forced to belong to a single community. Meanwhile, the result of the link community, gives a much more intuitive interpretation of the organization structure. For example, the links connected to the No.2 node are divided into two communities, blue (lower left) community and red (center) community, and the red links are connecting other red nodes while most of the blue links are connected to the blue nodes. The links connected to node No.1 or to other nodes that are located at the boundary of communities, show similar behavior. The link community scheme properly describes the multiple roles of the overlapping nodes, and it gives a more intuitive organization structure than the node community scheme, at least in this example.

Meanwhile, it is important to notice that the link community approach is not the perfect solution to the detection of the overlapping communities. For example, nodes Nos.9 and13 should belong to both the red community and blue community at the same time according to the result of link communities, but the connection between those two nodes is categorized only to the blue community. This result may not represent the relation between those two members properly because the interaction between those two members very likely would be related to both the red and blue communities, not being limited to only one community as the link community result suggests. Thus, exclusive partitioning of the links may not represent the community structure of network well when communities of links highly overlap. However, the link community approach is a reasonable approximation that is quite effective in the practical applications. Firstly, its computational complexity is of the same level of the node community approach, while most other methods of detecting overlapping communities~\cite{Nepusz2008,Gregory2011,Esquivel2011} require much more complex algorithms. Furthermore, the hard partitioning of links may not be an important issue if one is interested only in identifying the overlapping roles of the nodes because the degree of a node is usually larger than the number of the overlapping communities a node belongs to. For example, although the link between nodes Nos.9 and 13 is exclusively partitioned to the blue community, this result does not affect the detection of the overlapping roles of Nos.9 and 13.

\section{Community Results compared with metadata
\label{sec:RE1}}

The qualitative explanation of the community detection results, although interesting, has its limits in verifying the validity of the methods. A more solid approach would be comparing the community results with the metadata contained in the system, like the analysis in Ref.~\cite{YYAhn2009}. We analyzed four networks with rich metadata, which are listed in Table~\ref{tab:metadata-network}. The first is a sampled citation network of APS journal articles, which is constructed from the APS Data Sets for Research~\cite{aps-citation}. The sampled articles are the first- and second-level neighbors of a review paper~\cite{Albert2002} for complex networks. The metadata used to compare the results are the PACS (Physics and Astronomy Classification Scheme) numbers annotated to each article. Since the authors carefully choose the PACS numbers to make their articles well publicized, it is reasonable to consider the PACS numbers as rich and trustful metadata. The other three networks were previously constructed and analyzed in Ref.~\cite{YYAhn2009}. The metabolic network is constructed from \textit{E. coli} K-12 MG1655 strain, and the metadata used are the pathway annotations from the KEGG database~\cite{Kanehisa2000}. The philosopher network is a network of Wikipedia pages for philosophers, with each link representing the hyperlinks in the articles, and the metadata are the categories that each page belongs to. The last network analyzed is the word association network, which is constructed from the datasets about free association of word pairs~\cite{Nelson2004}, and the metadata are the meanings or definitions assigned to each word in WordNet database~\cite{Fellbaum1998}.

\begin{table}
\caption{\label{tab:metadata-network} The real-world networks with metadata. $N$ is the number of nodes, $M$ is the number of links, $C_\textrm{metadata}$ is the number of categories in the metadata, $C_\textrm{linkcom}$ is the number of communities detected by our method, and $C_\textrm{LC}$ is the number of communities detected by the link clustering method~\cite{YYAhn2009}.}
\begin{ruledtabular}
\begin{tabular}{cccccc}
 & $N$ & $M$ & $C_\textrm{metadata}$ &$C_\textrm{linkcom}$ & $C_\textrm{LC}$ \\
\hline
APS sample & 4755 & 29669 & 1076 & 339 & 14891\\
Metabolic~\cite{YYAhn2009} & 1042 & 17512 & 169 & 156 & 2304\\
Philosopher~\cite{YYAhn2009} & 1219 & 5972  & 5417 & 152 & 2777\\
Word Assoc.~\cite{YYAhn2009} & 5018 & 55232 & 13141 & 765 & 36654\\
\end{tabular}
\end{ruledtabular}
\end{table}

In these networks, each node is annotated with single or multiple metadata, and the metadata can be considered as the overlapping communities because they are closely related to the grouping of nodes. Also, the result of our method, in which the communities of links are detected, could be considered as the overlapping communities of nodes. Thus, comparing the result of our method with the pre-assigned metadata can be considered as comparing two different results of overlapping communities. Although several criterions have been proposed for comparing overlapping communities, none of them is as conclusive as the variation of information (VI)~\cite{Meila2006}, which is a well-defined and widely accepted criterion for comparing two \textit{non-overlapping} community partitions. In order to overcome the disadvantage of individual criterion for overlapping community, we compare the metadata with the link community results by two fundamentally different criterions, the extended normalized mutual information (NMI)~\cite{Lancichinetti2008} and the extended Jaccard index~\cite{Campello2010}, in order to observe the results from different aspects. Another extension of the mutual information for comparison of overlapping communities can be found in Ref.~\cite{Esquivel2011}. Although this method is a better approach compared to the extended NMI we used, it is not used in this work because in some of our examples one metadata may fully contain another metadata and the method cannot be used in this kind of cases.

\begin{figure}
\includegraphics[width=0.9\columnwidth]{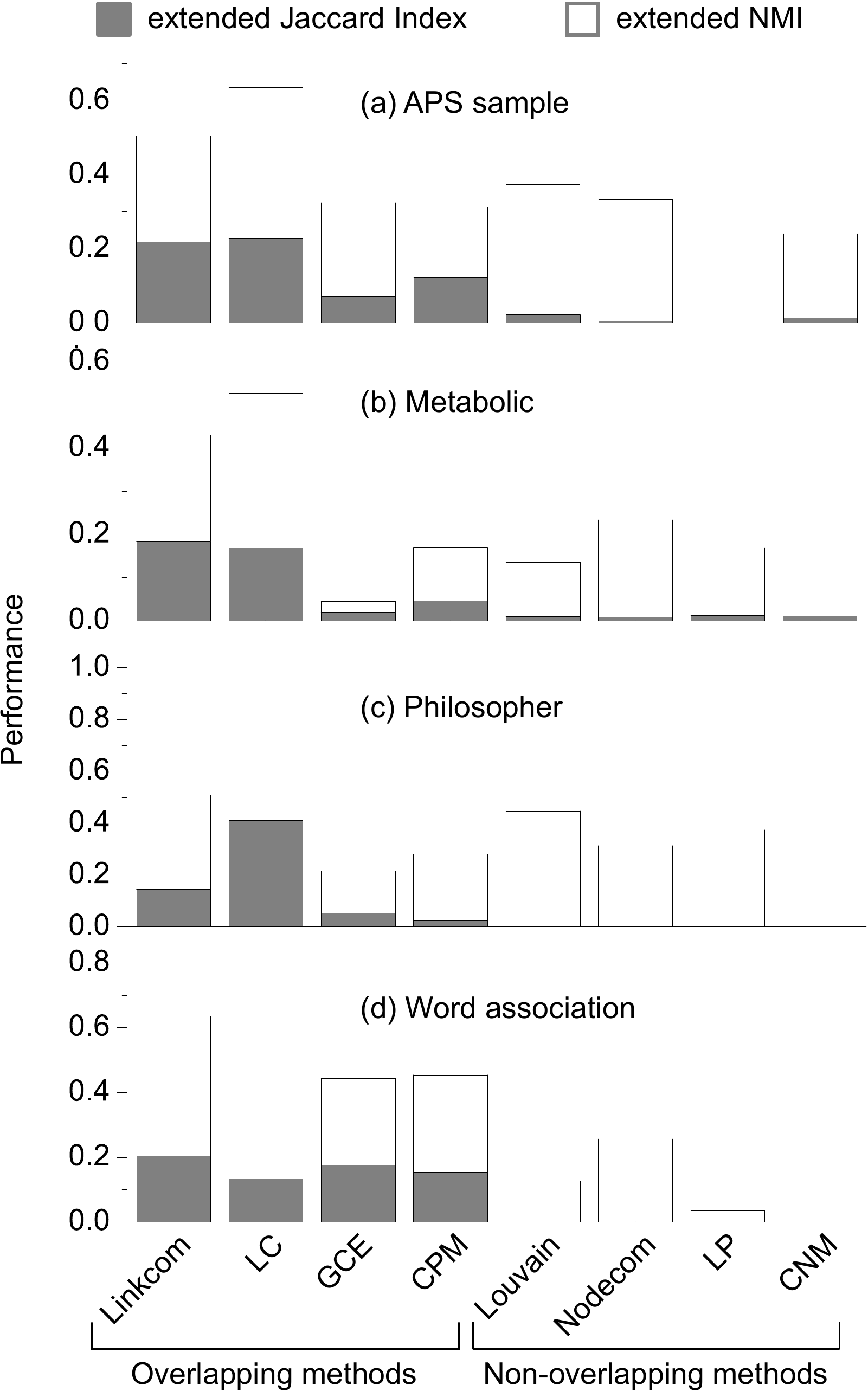}
\caption{\label{fig:performance} The performance test of various community-detecting methods. The communities detected by each method are compared with the metadata, and the performance is measured by the extended Jaccard index and extended NMI. Linkcom represents the result of our method, LC represents the link clustering method~\cite{YYAhn2009}, GCE represents the greedy clique expansion method~\cite{CLee2010}, CPM represents the clique percolation method, Louvain represents the fast unfolding method in Ref.~\cite{Blondel2008}, Nodecom represents the original map equation method~\cite{Rosvall2007}, LP represents the labal propagation method~\cite{Raghavan2007}, and CNM represents the Clauset-Newman-Moore method~\cite{Clauset2004a}. The first four methods are able to detect overlapping communities, and the last four methods are not~\cite{note1}.}
\end{figure}

The extended NMI is an information theory based measurement and is defined as
\begin{equation}
N(\bm{X}|\bm{Y}) = 1 - \frac{1}{2}[H(\bm{X}|\bm{Y})_\textrm{norm} + H(\bm{Y}|\bm{X})_\textrm{norm}],
\label{eq:norm_MI}
\end{equation}
where $\bm{X}$ and $\bm{Y}$ are two different partitions of overlapping communities and $H(\bm{X|Y})$ is the conditional entropy that measures the amount of information needed to infer $\bm{X}$ given the partition $\bm{Y}$. The extended NMI ranges from $0$ to $1$ and it equals to $1$ only when two partitions $\bm{X}$ and $\bm{Y}$ are identical. Meanwhile, the extended Jaccard coefficient falls into the category of external indexes that measure the similarity of two partitions statistically. This index is defined as
\begin{equation}
\omega(\bm{X,Y}) = \frac{a_G}{a_G + d_G,}
\label{eq:ext_jaccard}
\end{equation}
where $a_G$ and $d_G$ measure the agreement and disagreement of partition $\bm{X}$ and $\bm{Y}$ respectively. The index satisfies $\omega(\bm{X,Y}) \in [0,1]$, reaching $1$ only when $\bm{X}$ and $\bm{Y}$ are identical, and it reduces to the original Jaccard index in Ref.~\cite{Jain1988} when $\bm{X}$ and $\bm{Y}$ are non-overlapping partitions.

We applied our methods to the four networks and the detected communities are compared with the metadata by the extended Jaccard index and the extended NMI. The result is presented in Figure~\ref{fig:performance}, and the results of other community detection methods are also presented together to make a comparison. The first four methods, which are able to detect overlapping communities, show much better performance compared to the last four methods, which are able only to detect hard-partitioning communities. This result indicates the importance of detecting overlapping communities in recovering the properties of individual nodes. The first two methods, our method and link clustering method~\cite{YYAhn2009}, which are detecting overlapping communities for nodes by detecting link communities, show significantly better performance---both the extended Jaccard index and the extended NMI showing meaningfully large value through the four networks analyzed---compared to other methods, indicating the overlapping communities for nodes can be efficiently detected by finding the link communities. 

It is important to notice about our method and link clustering method, that both detect link communities but detect the communities at different hierarchical scales. As listed in Table~\ref{tab:metadata-network}, the number of communities detected by the link clustering method is much larger than our method, indicating our method detects communities of relatively larger size and the link clustering method detects communities of relatively smaller size. It would be necessary to consider this scale factor when deciding which method to use in order to analyze the community structure of networks. It seems like this difference originates from the different optimization goal of two methods, but the fundamental cause of this difference is left unknown at this time.

\section{Comparison of link community and node  community}
\label{sec:RE2}

\begin{figure}
\includegraphics[width=0.9\columnwidth]{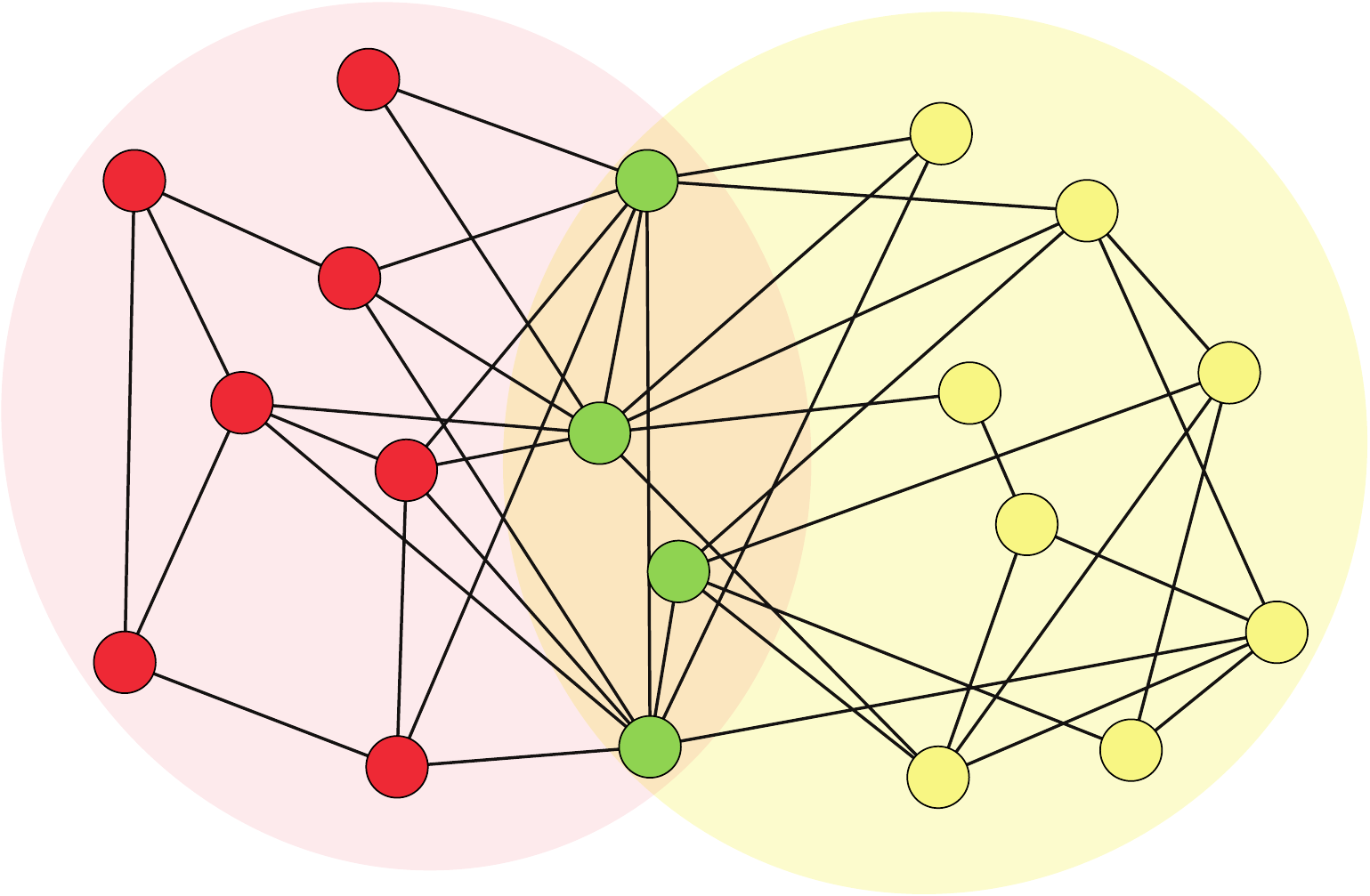}
\caption{\label{fig:model} (Color online) The model network that is proposed to verify the significance of overlap $\mathcal{O}$. This network is a variation of the Erd\"os-R\'enyi random network, and two communities, the red (left) and the blue (right), are embedded in the network. There are a total of $2N$ nodes in the network, and $2n$ of them (green or middle-gray nodes) are overlapping nodes while the other nodes are non-overlapping nodes, with $N-n$ nodes exclusively assigned to each community. The probability of connecting nodes in the same community is $p_\textrm{in}$, and it is much larger than $p_\textrm{out}$, which is the probability of connecting nodes from different communities. }
\end{figure}

\begin{figure}
\includegraphics[width=0.9\columnwidth]{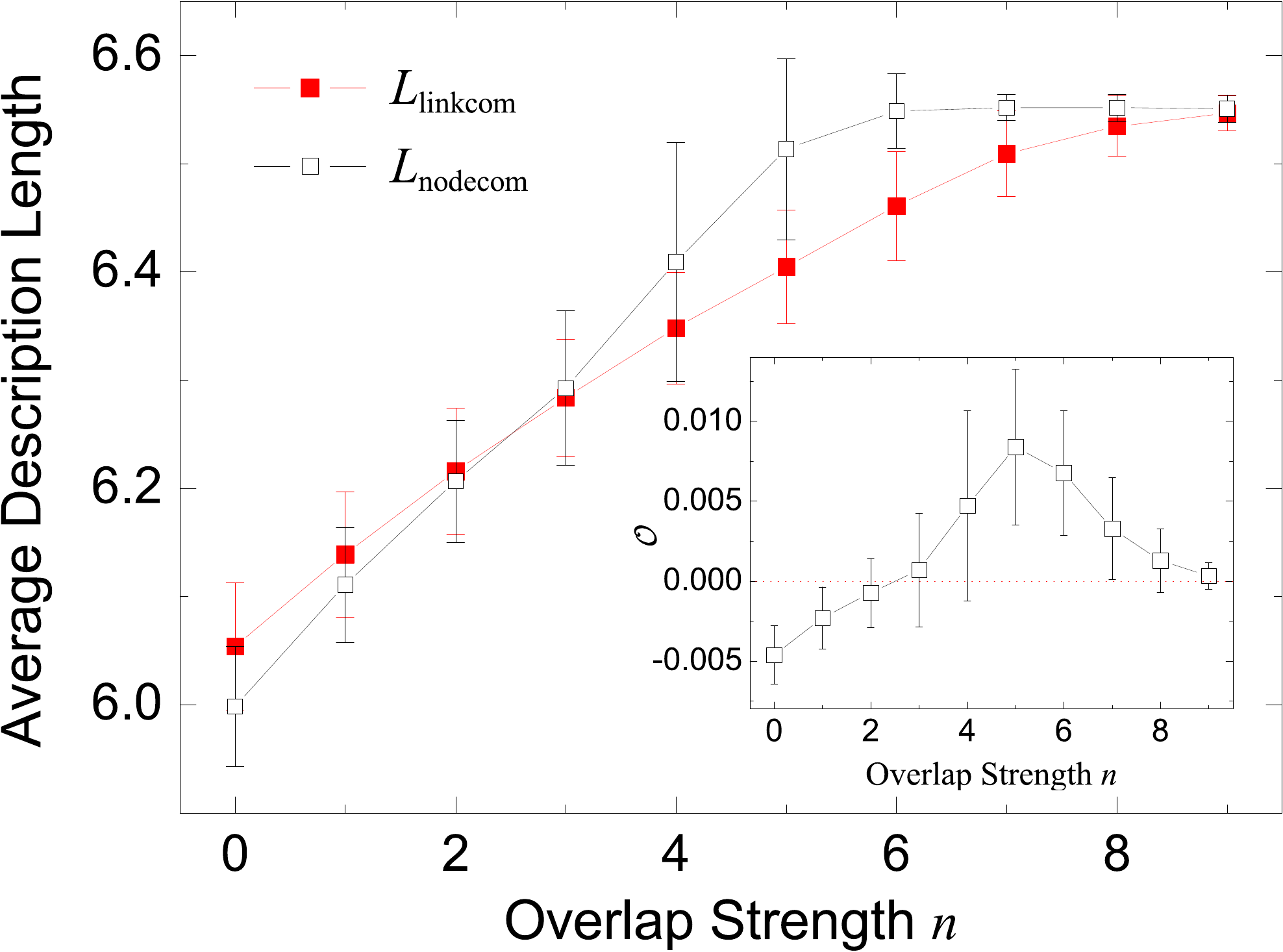}
\caption{\label{fig:compare} (Color online) The minimum average description length detected by our method and the original map equation method in the model network. The red (filled) squares represent the minimum of $L_\textrm{linkcom}$ given by our method, and the black (empty) squares represent the minimum of $L_\textrm{nodecom}$ given by the original map equation method, for different values of overlap strength $n$. The inset shows the value of the significance of overlap $\mathcal{O}$.}
\end{figure}

\begin{table*}
\caption{\label{tab:sig_overlap} The significance of overlap $\mathcal{O}$ measured in some real-world networks. The networks are listed by the descending order of $\mathcal{O}$.}
\begin{ruledtabular}
\begin{tabular}{ccccccc}
Network & $N$ & $M$ & $ <k>$ & $L_\textrm{linkcom}$ & $L_\textrm{nodecom}$ & $\mathcal{O}$\\
\hline
High-energy theory collaborations~\cite{Newman2001c} & 8361 & 15751 & 3.8  & 5.89  & 6.56  & 0.0539 \\
Network science collaborations~\cite{Newman2006a} & 1589 & 2742 & 3.5  & 3.48  & 3.77  & 0.0397\\ 
Power grid\cite{Watts1998} & 4941 & 6594 & 2.7  & 5.19  & 5.60  & 0.0380 \\
Amazon.com co-purchase~\cite{YYAhn2009} & 18142 & 46166 & 5.1  & 5.84  & 6.11  & 0.0224\\ 
Political blogs~\cite{Adamic2005} & 1490 & 19090 & 25.6  & 8.65  & 8.93  & 0.0163 \\
Word association~\cite{YYAhn2009} & 5018 & 55232 & 22.0  & 11.00  & 11.18  & 0.0080 \\
APS journal citations (sampled)~\cite{aps-citation} & 4755 & 29669 & 12.5  & 8.82  & 8.96  & 0.0076 \\
Protein-protein interaction~\cite{YYAhn2009} & 2729 & 12174 & 8.9  & 6.70  & 6.79  & 0.0068 \\
Word adjacencies~\cite{Newman2006a} & 112 & 425 & 7.6  & 6.27  & 6.35  & 0.0068 \\
\textit{Les Miserables}~\cite{Knuth1993} & 77 & 254 & 6.6  & 4.64  & 4.68  & 0.0043 \\
Political books~\cite{orgnet} & 105 & 441 & 8.4  & 5.44  & 5.48  & 0.0037 \\
\textbf{Zachary's karate club}~\cite{Zachary1977} & 34 & 78 & 4.6 & 4.28 & 4.31 & 0.0035\\
Dolphin social network~\cite{Lusseau2003} & 62 & 159 & 5.1  & 4.83  & 4.85  & 0.0024 \\
Philosopher~\cite{YYAhn2009} & 1219 & 5972 & 9.8  & 8.43  & 8.46  & 0.0018 \\
Jazz musicians collaborations~\cite{Gleiser2003} & 198 & 5484 & 55.4  & 6.91  & 6.91  & 0.0002 \\
\hline
\textit{C. Elegans} neural~\cite{Watts1998} & 297 & 2359 & 15.9  & 7.52  & 7.46  & -0.0041\\ 
\textit{E. coli} metabolic~\cite{YYAhn2009} & 1042 & 17512 & 33.6  & 8.33  & 8.25  & -0.0053 \\
American College football~\cite{Girvan2002} & 115 & 616 & 10.7  & 5.66  & 5.44  & -0.0199 \\
\end{tabular}
\end{ruledtabular}
\end{table*}

It is interesting to notice that in the result of karate club network, which is illustrated in Figure~\ref{fig:karate}, the map equation for link community, $L_\textrm{linkcom}$, is smaller than the map equation for node community, $L_\textrm{nodecom}$. Reminding that the map equation measures the amount of unknown information about the structure of the network assuming the community structure is already known, for each of $L_\textrm{linkcom}$ and $L_\textrm{nodecom}$ a smaller value of the map equation indicates that the community structure we assumed is a more proper description about the organization structure of the network. This reasoning can be extended to the comparison between $L_\textrm{linkcom}$ and $L_\textrm{nodecom}$. In both of the methods, the rule for random walk is the same, the second level codes are all assigned to the nodes, and the description length is measured in the same unit. Therefore, the only possible cause for the difference between $L_\textrm{linkcom}$ and $L_\textrm{nodecom}$ is the different rules for the first level codes, and this difference can be used to test which encoding rule is better---the link community or the node community. For example, if the minimum value of $L_\textrm{linkcom}$ is smaller than the minimum of $L_\textrm{nodecom}$, one can conclude that the link community scheme is better than the node community scheme in representing the organization structure of the network, because the link community scheme subtracted more information about the structure and left less unknown information in the path description. Instead, if the $L_\textrm{nodecom}$ is smaller, this means that there is no much overlap of communities over the nodes and the non-overlapping methods are good enough to study the community structure of the network.

To quantitively analyze the difference between $L_\textrm{nodecom}$ and $L_\textrm{linkcom}$, we propose a quantity called the \textit{significance of overlap}:
\begin{equation}
\mathcal{O} = \frac{ L_\textrm{nodecom} - L_\textrm{linkcom} }{ L_\textrm{nodecom} + L_\textrm{linkcom}}.
\end{equation}
This quantity measures how much better the link community scheme is compared to the node community scheme, and furthermore it can also be used to measure the overlapping strength of communities. The significance of overlap satisfies $\mathcal{O} \in (-1, 1)$, and it is positive when the link community scheme is better, being negative otherwise.

In order to check the validity of this quantity, we propose a model network (Figure~\ref{fig:model}) generated as follows. The model network is based on the Erd\"os-R\'enyi network~\cite{Erdos1959}, and two overlapping communities are embedded on the network. Among the $2N$ nodes of the network, $2n$ nodes are overlapping nodes, while $N-n$ nodes are exclusively assigned to each community. The probability of linking two nodes from the same community is $p_\textrm{in}$, and the probability of linking two nodes from different communities is $p_\textrm{out}$. The two communities overlaps more when $n$ is larger and overlaps less when $n$ is smaller. Therefore, $n$ can be considered as the parameter that controls the overlap strength of the two communities. Figure~\ref{fig:compare} shows the results of $L_\textrm{nodecom}$, $L_\textrm{linkcom}$ and the significance of overlap $\mathcal{O}$ for different values of $n$, while the set of parameters are fixed as $N = 50$, $<k> = 10$, $p_{out} / p_{in} = 15$. The error bar indicates the standard deviation over four hundred ensembles of the network realizations. When the overlap strength $n$ is small, $L_\textrm{nodecom}$ is much smaller than $L_\textrm{linkcom}$, indicating the node community scheme is better, and the significance of overlap $\mathcal{O}$ gets a negative value. As $n$ grows, the significance of overlap $\mathcal{O}$ gets larger and it starts to get a positive value, which means the link community scheme is better in describing the organization structure. When $n$ gets even larger, the overlap is too strong and the network is recognized as one community in both of the methods. Thus, the value of $\mathcal{O}$ falls to zero. This result matches our prediction well, therefore, the significance of overlap could be used as the quantitive measure of the strength of overlap.

We measured the significance of overlap $\mathcal{O}$ for some real-world networks and the results are listed in Table~\ref{tab:sig_overlap}. Although we do not fully understand how to interpret the exact value of $\mathcal{O}$ yet, some conclusions can still be made by comparing the values of $\mathcal{O}$ with the result of the karate club network (Figure~\ref{fig:karate}), in which the overlap of communities is well observed. The significance of overlap in the karate club network is $0.0035$, and many networks show a larger value of $\mathcal{O}$ than the karate club network. Many social networks---such as the collaboration networks of scientists, the network of political blogs, the social network in \textit{Les Miserables}, and the dolphin social network---show much stronger or similar degrees of community overlap compared to the karate club network, in accordance with the well-accepted knowledge that social communities tend to overlap with each other. The biological networks such as the \textit{C. elegans} neural network and the metabolic network show negative values of $\mathcal{O}$, which means the communities in these networks do not overlap much, while the protein-protein interaction network shows a positive value of $\mathcal{O}$. This result might be related to the different biological functions of the communities in these biological networks, and further investigations would be necessary. The college football network, in which the teams are divided into regional leagues and most games happened within the leagues, shows a non-overlapping community structure and this result strengthens the validity of the significance of overlap. Finally, the fact that many networks have positive values of $\mathcal{O}$ indicates the overlapping community structure exists in many real-world networks, and it is important to study the organization structure of these networks by detecting the overlapping communities, instead of insisting on the non-overlapping communities.

\section{Summary}
We proposed a method to detect link communities in networks by modifying the map equation method, which detects communities by minimizing the average description length of the random walk. In our method, the communities are assigned to links instead of nodes, the encoding rule for the random walk is modified to represent this change in the community structure, and the corresponding map equation for the link community is proposed. The map equation for link community could be used to detect the link communities by finding the link partitioning that gives the minimum value of the map equation, just like other quality functions, and most of the algorithms that were developed to maximize (or minimize) other quality functions could be used after minor modifications. 

One of the advantages of our methods is that the overlapping communities of nodes could be detected relatively easily, by defining the community of nodes by the communities of the links that are connected to the node. We tested our methods on some real-world networks by comparing the community results with the metadata of the nodes, and the result is compared with other community detection methods. The result shows that the communities detected by our method agree well with the metadata of the nodes, and the link community scheme is an efficient way to detect the overlapping communities of nodes.

Another important advantage of our methods is that the node community scheme and the link community scheme could be compared quantitatively. Since the difference between the map equation for the link community and the map equation for the node community comes only from the difference in community structure---the communities being assigned to the links or the nodes, the difference can be used to test which scheme, the link community or the node community, is better to represent the organization structure of the network. We used a quantity named as the significance of overlap to measure this difference in map equations, and the analysis of the significance of overlap in some real-world networks shows that many of the real-world networks can be better studied by the link community. Therefore, detecting the overlapping communities is necessary to understand the organization structure of the networks better, and finding link communities is an efficient way to detect the overlapping communities of nodes.

\begin{acknowledgments}
The authors thank Yong-Yeol Ahn and James P. Bagrow for providing us the network data along with the community results based on various community detecting methods. This work was supported by the Korean Systems Biology Research Project (20110002149) of the Ministry of Education, Science and Technology (MEST) through the National Research Foundation of Korea and by NAP of the Korean Research Council of Fundamental Science \& Technology(KRCF).
\end{acknowledgments}

%\bibliographystyle{apsrev}
%\bibliography{linkcom.bib}

\begin{thebibliography}{41}
\bibitem[]{note1}
The result of APS sample by LP method is missing from the figure, because the code for LP was not available.
\end{thebibliography}

\end{document}